\documentclass[12pt]{article}
\usepackage{amsmath}
\usepackage{amssymb}
\usepackage{amsfonts}

\renewcommand{\theequation}{\arabic{section}.\arabic{equation}}
\newcommand{\field}[1]{\mathbb{#1}}
\newcommand{\R}{\field{R}}
\DeclareMathOperator{\Res}{Res}

\title{QUASI-EXACTLY SOLVABLE SCHR\"ODINGER EQUATIONS, SYMMETRIC POLYNOMIALS, AND FUNCTIONAL BETHE ANSATZ METHOD}
\author{CHRISTIANE QUESNE\\ 
{\small\sl Physique Nucl\'eaire Th\'eorique et Physique Math\'ematique,  Universit\'e Libre de Bruxelles,} \\ 
{\small\sl Campus de la Plaine CP229, Boulevard~du Triomphe, B-1050 Brussels, Belgium}\\
{\small\sl cquesne@ulb.ac.be}}
\date{ }
\begin{document}
\baselineskip=22pt plus 1pt minus 1pt
\maketitle
\begin{abstract}
For applications to quasi-exactly solvable Schr\"odinger equations in quantum mechanics, we consider the general conditions that have to be satisfied by the coefficients of a second-order differential equation with at most $k+1$ singular points in order that this equation has particular solutions that are $n$th-degree polynomials. In a first approach, we show that such conditions involve $k-2$ integration constants, which satisfy a system of linear equations whose coefficients can be written in terms of elementary symmetric polynomials in the polynomial solution roots whenver such roots are all real and distinct. In a second approach, we consider the functional Bethe ansatz method in its most general form under the same assumption. Comparing the two approaches, we prove that the above-mentioned $k-2$ integration constants can be expressed as linear combinations of monomial symmetric polynomials in the roots, associated with partitions into no more than two parts. We illustrate these results by considering a quasi-exactly solvable extension of the Mathews-Lakshmanan nonlinear oscillator corresponding to $k=4$. 
\end{abstract}


%
%
\newpage
\section{INTRODUCTION}

In quantum mechanics, solving the Schr\"odinger equation is a fundamental problem for understanding physical systems. Exact solutions may be very useful for developing a constructive perturbation theory or for suggesting trial functions in variational calculus for more complicated cases. However, very few potentials can actually be exactly solved (see, {\it e.g.}, one of their lists in \cite{cooper}).\footnote{Here we do not plan to discuss the recent development of the exceptional orthogonal polynomials and the associated polynomially solvable analytic potentials (see, {\it e.g.}, \cite{gomez14} and references quoted therein).} These potentials are connected with second-order differential equations of hypergeometric type and their wavefunctions can be constructed by using the theory of corresponding orthogonal polynomials \cite{szego}.\par
%
%
A second category of exact solutions belongs to the so-called quasi-exactly solvable (QES) Schr\"odinger equations. These occupy an intermediate place between exactly solvable (ES) and non-solvable ones in the sense that only a finite number of eigenstates can be found explicitly by algebraic means, while the remaining ones remain unknown. The simplest QES problems, discovered in the 1980s, are characterized by a hidden sl(2,$\R$) algebraic structure \cite{turbiner87, turbiner88, ushveridze, gonzalez, turbiner16} and are connected with polynomial solutions of the Heun equation \cite{ronveaux}. Generalizations of this equation are related through their polynomial solutions to more complicated QES problems. Several procedures are employed in this context, such as the use of high-order recursion relations (see, {\it e.g.}, \cite{ciftci}) or the functional Bethe ansatz (FBA) method \cite{gaudin, ho, zhang}, which has proven very effective \cite{agboola12, agboola13, agboola14, cq17}.\par
%
%
The purpose of the present paper is to reconsider the general conditions under which a second-order differential equation $X(z) y''(z) + Y(z) y'(z) + Z(z) y(z) = 0$ with polynomial coefficients $X(z)$, $Y(z)$, and $Z(z)$ of respective degrees $k$, $k-1$, and $k-2$, has $n$th-degree polynomial solutions $y_n(z)$. Choosing appropriately the polynomial $Z(z)$ for such a purpose is known as the classical Heine-Stieltjes problem \cite{heine, stieltjes}. Such a differential equation with at most $k+1$ singular points covers the cases of the hypergeometric equation (for $k=2$), the Heun equation (for $k=3$), and generalized Heun equations (for $k\ge 4$), and plays therefore a crucial part in ES and QES quantum problems.\par
%
%
Here we plan to emphasize the key role played by symmetric polynomials in the polynomial solution roots whenever such roots are all real and distinct. This will be done by comparing two different approaches: a first one expressing the polynomial $Z(z)$ in terms of $k-2$ integration constants, and a second one based on the FBA method.\par
%
%
In Section 2, the problem of second-order differential equations with polynomial solutions is discussed and $k-2$ integration constants are introduced. In Section 3, the FBA method is derived in its most general form. A comparison between the two approaches is carried out in Section 4. The results so obtained are illustrated in Section 5 by considering a QES extension of the Mathews-Lakshmanan nonlinear oscillator. Finally, Section 6 contains the conclusion.\par
%
%
\section{SECOND-ORDER DIFFERENTIAL EQUATIONS WITH POLYNOMIAL SOLUTIONS AND INTEGRATION CONSTANTS}

On starting from a Schr\"odinger equation $(T+V)\psi(x) = E\psi(x)$, where the real variable $x$ varies in a given domain, the potential energy $V$ is a function of $x$, and the kinetic energy $T$ depends on $d/dx$ (and possibly on $x$ whenever the space is curved or the mass depends on the position \cite{cq04}), and on making an appropriate gauge transformation and a change of variable $x = x(z)$, one may arrive at a second-order differential equation with at most $k+1$ singular points,
\begin{equation}
  X(z) y''(z) + Y(z) y'(z) + Z(z) y(z) = 0,  \label{eq:red-eq-bis}
\end{equation}
where
\begin{equation}
  X(z) = \sum_{l=0}^k a_l z^l, \qquad Y(z) = \sum_{l=0}^{k-1} b_l z^l, \qquad Z(z) = \sum_{l=0}^{k-2} c_l z^l,
  \label{eq:X-Y-Z}
\end{equation}
and $a_l$, $b_l$, $c_l$ are some (real) constants. Polynomial solutions $y_n(z)$ of this equation will yield exact solutions $\psi_n(x)$ of the starting Schr\"odinger equation provided the latter are normalizable.\par
%
%
On deriving equation (\ref{eq:red-eq-bis}) $k-2$ times, we obtain
\begin{align}
  & X y^{(k)} + \left[\binom{k-2}{1} X' + Y\right] y^{(k-1)} + \left[\binom{k-2}{2} X'' + \binom{k-2}{1}
       Y' + Z \right] y^{(k-2)} + \cdots \nonumber \\
  & + \left[X^{(k-2)} + \binom{k-2}{1} Y^{(k-3)} + \binom{k-2}{2} Z^{(k-4)}\right] y'' + \left[Y^{(k-2)} +
       \binom{k-2}{1} Z^{(k-3)}\right] y' \nonumber \\
  & + Z^{(k-2)} y = 0,  \label{eq:diff-red-eq} 
\end{align}
which is a $k$th-order homogeneous differential equation with polynomial coefficients of degree not exceeding the corresponding order of differentiation. Since all its derivatives will have the same property, it can be differentiated $n$ times by using the new representation $y^{(n)}(z) = v_n(z)$. In such a notation, equation (\ref{eq:diff-red-eq}) can be written as
\begin{align}
  & X v_0^{(k)} + \left[\binom{k-2}{1} X' + Y\right] v_0^{(k-1)} + \left[\binom{k-2}{2} X'' 
       + \binom{k-2}{1} Y' + Z \right] v_0^{(k-2)} + \cdots \nonumber \\
  & + \left[X^{(k-2)} + \binom{k-2}{1} Y^{(k-3)} + \binom{k-2}{2} Z^{(k-4)}\right] v_0'' + \left[Y^{(k-2)} +
       \binom{k-2}{1} Z^{(k-3)}\right] v_0' \nonumber \\
  & + Z^{(k-2)} v_0 = 0. 
\end{align}
Its $n$th derivative can be easily shown to be given by
\begin{align}
  & \sum_{l=0}^k \binom{n+k-2}{k-l} X^{(k-l)} v_n^{(l)} + \sum_{l=0}^{k-1} \binom{n+k-2}{k-l-1} Y^{(k-l-1)}
       v_n^{(l)} \nonumber \\
  & + \sum_{l=0}^{k-2} \binom{n+k-2}{k-l-2} Z^{(k-l-2)} v_n^{(l)} = 0.  \label{eq:gen-eq}
\end{align}
When the coefficient of $v_n$ in (\ref{eq:gen-eq}) is equal to zero, {\it i.e.},
\begin{equation}
  \binom{n+k-2}{k} X^{(k)} + \binom{n+k-2}{k-1} Y^{(k-1)} + \binom{n+k-2}{k-2} Z^{(k-2)} = 0,
  \label{eq:condition}
\end{equation}
there only remains derivatives of $v_n$ in the equation. It is then obvious that the latter has a particular solution for which $v_n = y^{(n)}$ is a constant or, in other words, there exists a particular solution $y(z) = y_n(z)$ of equation (\ref{eq:red-eq-bis}) that is a polynomial of degree $n$.\par
%
%
On integrating equation (\ref{eq:condition}) $k-2$ times, we find that this occurs whenever $Z(z) = Z_n(z)$ is given by
\begin{equation}
  Z_n(z) = - \frac{n(n-1)}{k(k-1)} X''(z) - \frac{n}{k-1} Y'(z) + \sum_{l=0}^{k-3} C_{k-l-2,n} \frac{z^l}{l!},
  \label{eq:h_n}
\end{equation}
where $C_{1,n}, C_{2,n}, \ldots, C_{k-2,n}$ are $k-2$ integration constants. This is the first main result of this paper.
\par
%
%
It is worth observing that in the hypergeometric case, we have $k=2$ so that the polynomial $Z(z)$ reduces to a constant $\lambda$. Then $\lambda = \lambda_n$, where, in accordance with equation (\ref{eq:h_n}),
\begin{equation}
  \lambda_n = - \tfrac{1}{2} n(n-1) X''(z) - n Y'(z),
\end{equation}
with no integration constant, which is a well-known result \cite{szego}. In the Heun-type equation case, we have $k=3$ and the linear polynomial $Z(z) = Z_n(z)$ is given by
\begin{equation}
  Z_n(z) = - \tfrac{1}{6} n(n-1) X''(z) - \tfrac{1}{2} n Y'(z) + C_n
\end{equation}
in terms of a single integration constant $C_n$, which is a result previously derived by Karayer, Demirhan, and B\"uy\"ukk\i l\i \c c \cite{karayer15a}.\par
%
%
On inserting now equation (\ref{eq:X-Y-Z}) in (\ref{eq:h_n}) and equating the coefficients of equal powers of $z$ on both sides, we obtain the set of relations
\begin{align}
  c_{k-2} &= - n(n-1) a_k - n b_{k-1}, \label{eq:rel-1} \\
  c_l &= - \frac{n(n-1)}{k(k-1)} (l+2)(l+1) a_{l+2} - \frac{n}{k-1} (l+1) b_{l+1} + \frac{C_{k-l-2,n}}{l!}, \nonumber \\
  & \qquad l=0, 1, \ldots, k-3.  \label{eq:rel-2}
\end{align}
\par
%
%
The $n$th-degree polynomial solutions $y_n(z)$ of equation (\ref{eq:red-eq-bis}) can be written as
\begin{equation}
  y_n(z) = \prod_{i=1}^n (z-z_i),  \label{eq:roots}
\end{equation}
where from now on we assume that the roots $z_1, z_2, \ldots, z_n$ are real and distinct. We now plan to show that the integration constants $C_{1,n}, C_{2,n}, \ldots, C_{k-2,n}$ satisfy a system of linear equations whose coefficients can be expressed in terms of elementary symmetric polynomials in $z_1, z_2, \ldots, z_n$ \cite{littlewood},
\begin{equation}
\begin{split}
  e_l &\equiv e_l(z_1, z_2, \ldots, z_n) = \sum_{1\le i_1< i_2< \cdots < i_l\le n} z_{i_1} z_{i_2} \ldots z_{i_l}, \qquad 
      l=1, 2, \ldots, n, \\
  e_0 &\equiv 1.
\end{split}  \label{eq:e_l}
\end{equation}
\par
%
%
We can indeed rewrite $y_n(z)$ in (\ref{eq:roots}) as
\begin{equation}
  y_n(z) = \sum_{m=0}^n (-1)^{n-m} e_{n-m} z^m,
\end{equation}
so that
\begin{align}
  y'_n(z) &= \sum_{m=0}^{n-1} (-1)^{n-m-1} (m+1) e_{n-m-1} z^m, \\
  y''_n(z) &= \sum_{m=0}^{n-2} (-1)^{n-m-2} (m+2)(m+1) e_{n-m-2} z^m.
\end{align}
On inserting these expressions in equation (\ref{eq:red-eq-bis}) and taking equation (\ref{eq:X-Y-Z}) into account, we get
\begin{align}
  & \sum_{l=0}^k a_l z^l \sum_{m=0}^{n-2} (-1)^{n-m-2} (m+2)(m+1) e_{n-m-2} z^m \nonumber \\
  & + \sum_{l=0}^{k-1} b_l z^l \sum_{m=0}^{n-1} (-1)^{n-m-1} (m+1) e_{n-m-1} z^m \nonumber \\
  & + \sum_{l=0}^{k-2} c_l z^l \sum_{m=0}^n (-1)^{n-m} e_{n-m} z^m = 0.  \label{eq:red-eq-ter}
\end{align}
\par
%
%
Here $l+m$ runs from 0 to $k+n-2$. Let us therefore set $l+m = k+n-r$, where $r=2, 3, \ldots, k+n$. Equation (\ref{eq:red-eq-ter}) can then be rewritten as
\begin{align}
  & \sum_{r=2}^k \left\{\sum_{p=0}^{r-2} (-1)^p \left[a_{k-r+2+p} (n-p)(n-p-1) + b_{k-r+1+p} (n-p) + c_{k-r+p}\right]
       e_p\right\} z^{k+n-r} \nonumber \\
  & \qquad + \text{(lower-degree terms with $k+1 \le r \le k+n$)} = 0.
\end{align}
On setting to zero the coefficients of $z^{k+n-r}$, $r=2, 3, \ldots, k$, we obtain the relations
\begin{align}
  & \sum_{p=0}^{r-2} (-1)^p \left[a_{k-r+2+p} (n-p)(n-p-1) + b_{k-r+1+p} (n-p) + c_{k-r+p}\right] e_p = 0, \nonumber \\
  & \qquad r=2, 3, \ldots, k.  
\end{align}
For $r=2$, we simply get $n(n-1) a_k + n b_{k-1} + c_{k-2} = 0$, which is automatically satisfied due to equation (\ref{eq:rel-1}). We are therefore left with the $k-2$ relations
\begin{align}
  & \sum_{p=0}^{r-3} (-1)^p \left[a_{k-r+2+p} (n-p)(n-p-1) + b_{k-r+1+p} (n-p) + c_{k-r+p}\right] e_p \nonumber \\
  & + (-1)^{r-2} [a_k (n-r+2)(n-r+1) + b_{k-1} (n-r+2) + c_{k-2}] e_{r-2} = 0, \nonumber \\
  & \qquad r=3, 4, \ldots, k.
\end{align} 
\par
%
%
After substituting the right-hand sides of equations (\ref{eq:rel-1}) and (\ref{eq:rel-2}) for $c_{k-2}$ and $c_{k-r+p}$ in these relations, we obtain a system of $k-2$ linear equations for the $k-2$ integration constants $C_{1,n}, C_{2,n}, \ldots, C_{k-2,n}$, 
\begin{align}
 & \sum_{p=0}^{r-3} (-1)^p \frac{C_{r-2-p,n}}{(k-r+p)!} e_p = - \sum_{p=0}^{r-3} (-1)^p \biggl\{\frac{1}{k(k-1)} 
     \Bigl[(r-p-2)(2k-r+p+1) n^2 \nonumber \\
 &\quad - [2pk^2 + 2k(r-2p-2) - (r-p-2)(r-p-1)] n + k(k-1)p(p+1)\Bigr] a_{k-r+2+p} \nonumber \\
 &\quad + \frac{1}{k-1} [(r-p-2) n - (k-1)p] b_{k-r+1+p}\biggr\} e_p \nonumber \\
 &\quad - (-1)^{r-1} (r-2) [(2n-r+1) a_k + b_{k-1}] e_{r-2}, \qquad r=3, 4, \ldots, k,  \label{eq:syst-C}
\end{align}
whose coefficients are expressed in terms of elementary symmetric polynomials (\ref{eq:e_l}) in the roots of the polynomial solutions of equation (\ref{eq:red-eq-bis}). This is the second main result of this paper.
\par
%
%
The determinant of this system having zeros above the diagonal is easily determined to be given by $\left[\prod_{r=3}^k (k-r)!\right]^{-1} \ne 0$. It is therefore obvious that the constants $C_{1,n}, C_{2,n}, \ldots , C_{k-2,n}$ can be calculated successively from the equations corresponding to $r=3, 4, \ldots, k$.\par
%
%
It turns out that instead of elementary symmetric polynomials $e_l(z_1, z_2, \ldots, z_n)$, defined in equation (\ref{eq:e_l}), it is more appropriate to express the solution in terms of monomial symmetric polynomials in $z_1, z_2, \ldots, z_n$, 
\begin{equation}
  m_{(\lambda_1, \lambda_2, \ldots, \lambda_n)}(z_1, z_2, \ldots, z_n) = \sum_{\pi \in S_{\lambda}} 
  z_{\pi(1)}^{\lambda_1} z_{\pi(2)}^{\lambda_2} \ldots z_{\pi(n)}^{\lambda_n},  \label{eq:monomial}
\end{equation}
where $(\lambda_1, \lambda_2, \ldots, \lambda_n)$ denotes a partition and $S_{\lambda}$ is the set of permutations giving distinct terms in the sum \cite{littlewood}. The derivation of the first three constants $C_{1,n}$, $C_{2,n}$, and $C_{3,n}$ is outlined in Appendix. In particular, it is shown there that $m_{(1^3,\dot{0})}$ (a dot over zero meaning that it is repeated as often as necessary), corresponding to a partition into more than two parts and which in principle might appear in $C_{3,n}$, actually does not occur because it has a vanishing coefficient. This is a general property that we have observed for the first six constants that we have computed explicitly and which all agree with the general formula
\begin{align}
  \frac{C_{q,n}}{(k-2-q)!} &= - \sum_{t=0}^{q-1} [2(n-1) a_{k-t} + b_{k-t-1}] m_{(q-t,\dot{0})} - \sum_{s=1}^{[q/2]}
       \sum_{t=0}^{q-2s} 2a_{k-t}\, m_{(q-t-s,s,\dot{0})} \nonumber \\
  &\quad - \frac{n(n-1)}{k(k-1)} q(2k-q-1) a_{k-q} - \frac{n}{k-1} q b_{k-q-1}, \nonumber \\
  &\qquad q=1, 2, \ldots, k-2,  \label{eq:C}
\end{align}
where $[q/2]$ denotes the largest integer contained in $q/2$.\par
%
%
At this stage, equation (\ref{eq:C}) is a conjecture, which might be proved from (\ref{eq:syst-C}) by determining $C_{q,n}$ for any $q \in \{1, 2, \ldots, k-2\}$. This would, however, be a rather complicated derivation. In Section 4, we will proceed to show that a much easier proof of (\ref{eq:C}) can be found by comparing the results of the present approach with those of the FBA one.\par
%
%
\section{FUNCTIONAL BETHE ANSATZ METHOD}

\setcounter{equation}{0}

In its most general form, the FBA method also starts from equation (\ref{eq:red-eq-bis}), with $X(z)$, $Y(z)$, $Z(z)$ given in (\ref{eq:X-Y-Z}), and considers polynomial solutions of type (\ref{eq:roots}) with real and distinct roots $z_1, z_2, \ldots, z_n$ \cite{gaudin, ho, zhang}. Equation (\ref{eq:red-eq-bis}) is then rewritten as
\begin{equation}
  - c_0 = \left(\sum_{l=0}^k a_l z^l\right) \sum_{i=1}^n \frac{1}{z-z_i} \sum_{\substack{
  j=1 \\
  j\ne i}}^n \frac{2}{z_i-z_j} + \left(\sum_{l=0}^{k-1} b_l z^l\right) \sum_{i=1}^n \frac{1}{z-z_i} + \sum_{l=1}^{k-2}
  c_l z^l.  \label{eq:constant}
\end{equation}
The left-hand side of this equation is a constant, while the right-hand one is a meromorphic function with simple poles at $z=z_i$ and a singularity at $z=\infty$. Since the residues at the simple poles are given by
\begin{equation}
  \Res(-c_0)_{z=z_i} = \left(\sum_{l=0}^k a_l z_i^l\right) \sum_{\substack{
  j=1 \\
  j\ne i}}^n \frac{2}{z_i-z_j} + \sum_{l=0}^{k-1} b_l z_i^l, 
\end{equation}
equation (\ref{eq:constant}) yields
\begin{align}
  - c_0 &= \sum_{l=1}^k a_l \sum_{i=1}^n \sum_{m=0}^{l-1} z_i^m z^{l-m-1}\sum_{\substack{
      j=1 \\
      j\ne i}}^n \frac{2}{z_i-z_j} + \sum_{l=1}^{k-1} b_l \sum_{i=1}^n \sum_{m=0}^{l-1} z_i^m z^{l-m-1} \nonumber \\
  &\quad + \sum_{l=1}^{k-2} c_l z^l + \sum_{i=1}^n \frac{\Res(-c_0)_{z=z_i}}{z-z_i}.
\end{align}
On defining 
\begin{equation}
  S_m = \sum_{i=1}^n \sum_{\substack{
      j=1 \\
      j\ne i}}^n \frac{z_i^m}{z_i-z_j}, \qquad T_m = \sum_{i=1}^n z_i^m,
\end{equation}
and observing that $S_0=0$, this relation becomes
\begin{equation}
  - c_0 = 2 \sum_{l=2}^k a_l \sum_{m=1}^{l-1} S_m z^{l-m-1} + \sum_{l=1}^{k-1} b_l \sum_{m=0}^{l-1} T_m z^{l-m-1}
  + \sum_{l=1}^{k-2} c_l z^l + \sum_{i=1}^n \frac{\Res(-c_0)_{z=z_i}}{z-z_i}, 
\end{equation}
or, with $q \equiv l-m-1$ in the first two terms and $q \equiv l$ in the third one,
\begin{equation}
  - c_0 = 2 \sum_{q=0}^{k-2} z^q \sum_{m=1}^{k-1-q} a_{q+m+1} S_m + \sum_{q=0}^{k-2} z^q \sum_{m=0}^{k-2-q}
       b_{q+m+1} T_m + \sum_{q=1}^{k-2} z^q c_q + \sum_{i=1}^n \frac{\Res(-c_0)_{z=z_i}}{z-z_i}. 
\end{equation}
\par
%
%
The right-hand side of this equation will be a constant if and only if the coefficients of $z^q$, $q=1, 2, \ldots, k-2$, and all the residues at the simple poles vanish. This yields $c_q$, $q=1, 2, \ldots, k-2$, in terms of the coefficients of $X(z)$, $Y(z)$, and the roots of $y_n(z)$,
\begin{equation}
  c_q = - 2 \sum_{m=1}^{k-1-q} a_{q+m+1} S_m - \sum_{m=0}^{k-2-q} b_{q+m+1} T_m, \qquad q=1, 2, \ldots, k-2,
  \label{eq:c_q}
\end{equation}
as well as the $n$ algebraic equations determining the roots, i.e., the Bethe ansatz equations,
\begin{equation}
  \sum_{\substack{
  j=1 \\
  j\ne i}}^n \frac{2}{z_i-z_j} + \frac{\sum_{l=0}^{k-1} b_l z_i^l}{\sum_{l=0}^k a_l z_i^l} = 0, \qquad i=1, 2, \ldots, n.
\end{equation}
The remaining constant leads to the value of $c_0$,
\begin{equation}
  c_0 = - 2 \sum_{m=1}^{k-1} a_{m+1} S_m - \sum_{m=0}^{k-2} b_{m+1} T_m.  \label{eq:c_0}
\end{equation}
\par
%
%
It remains to find the explicit expressions of $S_m$ and $T_m$. For the smallest allowed $m$ values, it is obvious that
\begin{equation}
  S_1 = \tfrac{1}{2} n(n-1), \qquad T_0 = n.
\end{equation}
For higher $m$ values, $S_m$ can be written as a linear combination of monomial symmetric polynomials in $z_1, z_2, \ldots, z_n$. From
\begin{equation}
  S_m = \frac{1}{2} \sum_{i=1}^n \sum_{\substack{
  j=1 \\
  j\ne i}}^n \frac{z_i^m - z_j^m}{z_i-z_j} = \frac{1}{2} \sum_{i=1}^n \sum_{\substack{
  j=1 \\
  j\ne i}}^n \sum_{p=0}^{m-1} z_i^{m-1-p} z_j^p,
\end{equation}
we get for odd $m \ge 3$,
\begin{align}
  S_m &= \frac{1}{2} \sum_{i=1}^n \sum_{\substack{
      j=1 \\
      j\ne i}}^n \Biggl[z_i^{m-1} + z_j^{m-1} + \sum_{p=1}^{(m-3)/2} \bigl(z_i^{m-1-p} z_j^p + z_i^p z_j^{m-1-p}\bigr)
      + z_i^{(m-1)/2} z_j^{(m-1)/2}\Biggr] \nonumber \\
  &= (n-1) \sum_{i=1}^n z_i^{m-1} + \sum_{\substack{
      i, j=1 \\
      i \ne j}}^n \sum_{p=1}^{(m-3)/2} z_i^{m-1-p} z_j^p + \sum_{\substack{
      i, j=1 \\
      i < j}}^n z_i^{(m-1)/2} z_j^{(m-1)/2} \nonumber \\
  &= (n-1) m_{(m-1,\dot{0})} + \sum_{p=1}^{(m-1)/2} m_{(m-1-p,p,\dot{0})}, 
\end{align}
and for even $m \ge 2$,
\begin{align}
  S_m &= \frac{1}{2} \sum_{i=1}^n \sum_{\substack{
      j=1 \\
      j\ne i}}^n \Biggl[z_i^{m-1} + z_j^{m-1} + \sum_{p=1}^{(m-2)/2} \bigl(z_i^{m-1-p} z_j^p + z_i^p z_j^{m-1-p}\bigr)
      \Biggr] \nonumber \\
  &= (n-1) \sum_{i=1}^n z_i^{m-1} + \sum_{\substack{
      i, j=1 \\
      i \ne j}}^n \sum_{p=1}^{(m-2)/2} z_i^{m-1-p} z_j^p  \nonumber \\
  &= (n-1) m_{(m-1,\dot{0})} + \sum_{p=1}^{(m-2)/2} m_{(m-1-p,p,\dot{0})}. 
\end{align}
Hence, 
\begin{equation}
  S_m = (n-1) m_{(m-1,\dot{0})} + \sum_{p=1}^{[(m-1)/2]} m_{(m-1-p,p,\dot{0})}, \qquad m \ge 2.
\end{equation}
\par
%
%
{}Furthermore, it is obvious that
\begin{equation}
  T_m = m_{(m,\dot{0})}, \qquad m \ge 1.
\end{equation}
\par
%
%
On replacing $S_m$ and $T_m$ by their explicit values in (\ref{eq:c_q}) and (\ref{eq:c_0}), we get
\begin{align}
  c_{k-2} &= - n(n-1) a_k - n b_{k-1},  \label{eq:rel-1-bis} \\
  c_l &= - n(n-1) a_{l+2} - n b_{l+1} - 2 \sum_{m=2}^{k-1-l} a_{l+m+1} \biggl[(n-1) m_{(m-1,\dot{0})} \nonumber \\
  &\quad + \sum_{p=1}^{[(m-1)/2]} m_{(m-1-p,p,\dot{0})}\biggr] - \sum_{m=1}^{k-2-l} b_{l+m+1} m_{(m,\dot{0})},
       \nonumber \\
  &\qquad l=0, 1, \ldots, k-3,  \label{eq:rel-2-bis}
\end{align}
in terms of monomial symmetric polynomials in the polynomial roots. This is the third main result of this paper.\par
%
%
\section{COMPARISON BETWEEN THE TWO APPROACHES}

\setcounter{equation}{0}

Direct comparison between equations (\ref{eq:rel-1}), (\ref{eq:rel-2}) and equations (\ref{eq:rel-1-bis}), (\ref{eq:rel-2-bis}) shows that $c_{k-2}$ is given by the same expression in both approaches, while for $l=0, 1, \ldots, k-3$, $c_l$ is written in terms of $a_{l+2}$ and $b_{l+1}$, as well as the integration constant $C_{k-l-2,n}$ in the first one or a linear combination of monomial symmetric polynomials in $z_1, z_2, \ldots, z_n$ in the second one.\par
%
%
Equating the two expressions for $c_l$, $l=0, 1, \ldots, k-3$, yields
\begin{align}
  \frac{C_{k-2-l,n}}{l!} &= - 2 \sum_{m=2}^{k-1-l} a_{l+m+1} \biggl[(n-1) m_{(m-1,\dot{0})} + \sum_{p=1}^{[(m-1)/2]}
       m_{(m-1-p,p,\dot{0})}\biggr] \nonumber \\
  &\quad - \sum_{m=1}^{k-2-l} b_{l+m+1} m_{(m,\dot{0})} - \frac{n(n-1)}{k(k-1)} (k-l-2)(k+l+1) a_{l+2} \nonumber \\
  &\quad - \frac{n}{k-1} (k-l-2) b_{l+1}, \qquad l=0, 1, \ldots, k-3.  \label{eq:C-bis}
\end{align}
On setting $q=k-2-l$ in equation (\ref{eq:C-bis}), the latter becomes
\begin{align}
  \frac{C_{q,n}}{(k-2-q)!} &= - 2 \sum_{m=2}^{q+1} a_{k+m-1-q} \biggl[(n-1) m_{(m-1,\dot{0})} + \sum_{p=1}^{[(m-1)/2]}
       m_{(m-1-p,p,\dot{0})}\biggr] \nonumber \\
  &\quad - \sum_{m=1}^{q} b_{k+m-1-q} m_{(m,\dot{0})} - \frac{n(n-1)}{k(k-1)} q(2k-q-1) a_{k-q} \nonumber \\
  &\quad - \frac{n}{k-1} q b_{k-q-1}, \qquad l=0, 1, \ldots, k-3,  \label{eq:C-ter}
\end{align} 
where we see that the last two terms on the right-hand side coincide with the corresponding ones in equation (\ref{eq:C}). The other terms can also be easily converted into those of equation (\ref{eq:C}) by changing the summation indices. With $t=q+1-m$ and $t=q-m$, we can indeed rewrite
\begin{equation}
  \sum_{m=2}^{q+1} a_{k+m-1-q} m_{(m-1,\dot{0})} = \sum_{t=0}^{q-1} a_{k-t} m_{(q-t,\dot{0})}
\end{equation}
and
\begin{equation}
  \sum_{m=1}^q b_{k+m-1-q} m_{(m,\dot{0})} = \sum_{t=0}^{q-1} b_{k-t-1} m_{(q-t,\dot{0})},
\end{equation}
respectively. Furthermore, $t=q+1-m$ and $s=p$ lead to
\begin{align}
  & \sum_{m=2}^{q+1} a_{k+m-1-q} \sum_{p=1}^{[(m-1)/2]} m_{(m-1-p,p,\dot{0})} \nonumber \\
  &= \sum_{t=0}^{q-1} a_{k-t} \sum_{s=1}^{[(q-t)/2]} m_{(q-t-s,s,\dot{0})} \nonumber \\
  &= \sum_{s=1}^{[q/2]} \sum_{t=0}^{q-2s} a_{k-t} m_{(q-t-s,s,\dot{0})}.
\end{align}
\par
%
%
Collecting all the terms shows that equation (\ref{eq:C-ter}) coincides with equation (\ref{eq:C}), which is therefore proved. This is the fourth main result of the present paper.\par
%
%
\section{EXAMPLE: QES EXTENSION OF THE MATHEWS-LAKSHMANAN NONLINEAR OSCILLATOR}

\setcounter{equation}{0}

The purpose of the present section is to illustrate the results of previous ones by considering a QES extension of the Mathews-Lakshmanan nonlinear oscillator in the simplest case not amenable to an sl(2,$\R$) description, and which corresponds to $k=4$.\par
%
%
The quantum version of the classical Mathews-Lakshmanan nonlinear oscillator \cite{mathews} is described by the Hamiltonian \cite{carinena04, carinena07}
\begin{equation}
  H = - (1+\lambda x^2) \frac{d^2}{dx^2} - \lambda x \frac{d}{dx} + \frac{\beta(\beta+\lambda)x^2}
  {1+\lambda x^2},  \label{eq:M-L}
\end{equation}
where $\beta$ plays the role of the frequency $\omega$ in the standard oscillator and the nonlinearity parameter $\lambda\ne0$ enters both the potential energy term and the kinetic energy one, giving rise there to the position-dependent mass $m(x) = (1+\lambda x^2)^{-1}$. According to whether $\lambda>0$ or $\lambda<0$, the range of the coordinate $x$ is $(-\infty, \infty)$ or $(-1/\sqrt{|\lambda|}, 1/\sqrt{|\lambda|})$. Such a Hamiltonian is formally self-adjoint with respect to the measure $d\mu = (1+\lambda x^2)^{-1/2} dx$. The corresponding Schr\"odinger equation is ES \cite{carinena04, carinena07} and its bound-state wavefunctions can be expressed in terms of Gegenbauer polynomials \cite{schultze}.\par
%
%
Noting that the potential energy term in (\ref{eq:M-L}) can also be written as
\begin{equation}
  V_0(x) = \lambda A - \frac{\lambda A}{1+\lambda x^2}, \qquad \text{where} \quad A = \frac{\beta}{\lambda}
  \left(\frac{\beta}{\lambda}+1\right),
\end{equation}
let us extend it to
\begin{equation}
  V_m(x) = \lambda A - \frac{\lambda A}{1+\lambda x^2} + \lambda \sum_{k=1}^{2m} B_k (1+\lambda x^2)^k,
  \label{eq:V_m}
\end{equation}
where $m$ may take the values $m=1$, 2, 3,~\ldots, $A$, $B_1$, $B_2$, \ldots, $B_{2m}$ are $2m+1$ parameters, and the range of $x$ is the same as before. The starting Schr\"odinger equation therefore reads
\begin{equation}
  \left(- (1+\lambda x^2) \frac{d^2}{dx^2} - \lambda x \frac{d}{dx} + \lambda A - \frac{\lambda A}{1+\lambda x^2}
   + \lambda \sum_{k=1}^{2m} B_k (1+\lambda x^2)^k - E\right) \psi(x) = 0.  \label{eq:NLHO}
\end{equation}
To reduce it to an equation of type (\ref{eq:red-eq-bis}), let us make the change of variable
\begin{equation}
  z = \frac{1}{1+\lambda x^2}
\end{equation}
and the gauge transformation
\begin{equation}
  \psi(x) = x^p z^a \exp\left(- \sum_{j=1}^m \frac{b_j}{z^j}\right) y(z),
\end{equation}
where $p=0, 1$ is related to the parity $(-1)^p = +1, -1$, and $a$, $b_1$, $b_2$, \ldots, $b_m$ are $m+1$ parameters connected with the previous ones. It turns out that $k$ in (\ref{eq:X-Y-Z}) is related to $m$ in (\ref{eq:V_m}) by the relation $k = m+2$.\par
%
%
{}For $m=2$, for instance, equation (\ref{eq:NLHO}) yields
\begin{align}
  & \biggl\{-4z^3(1-z) \frac{d^2}{dz^2} + 2[(4a+3)z^3 - 2 (2a-2b_1+1-p)z^2 - 4(b_1-2b_2)z - 8b_2] \frac{d}{dz}
     \nonumber \\
  & + [2a(2a+1) - A]z^2 + [- 4a^2 + 8ab_1 + 4ap - 2b_1 - p - \epsilon]z + B_1 \nonumber \\
  & - 4b_1(2a-b_1-1-p) + 4b_2(4a-3) \biggr\} y(z) = 0,
\end{align} 
after setting $E = \lambda(\epsilon + A)$ and
\begin{equation}
  B_2 = 4[b_1^2 + 2b_2(2a-2b_1-2-p)], \qquad B_3 = 16b_2(b_1-b_2), \qquad B_4 = 16b_2^2.  \label{eq:B}
\end{equation}
\par
%
%
With the identifications
\begin{equation}
\begin{split}
  & a_4 \rightarrow 4, \qquad a_3 \rightarrow -4, \qquad a_2, a_1, a_0 \rightarrow 0, \\
  & b_3 \rightarrow 2(4a+3), \qquad b_2 \rightarrow -4 (2a-2b_1+1-p), \qquad b_1 \rightarrow -8 (b_1-2b_2), \\
  & \quad b_0 \rightarrow -16b_2, \\
  & c_2 \rightarrow 2a(2a+1) - A, \qquad c_1 \rightarrow - 4a^2 + 8ab_1 + 4ap - 2b_1 - p - \epsilon, \\
  & \quad c_0 \rightarrow B_1 - 4b_1(2a-b_1-1-p) + 4b_2(4a-3)
\end{split}
\end{equation}
in equation (\ref{eq:X-Y-Z}), from equations (\ref{eq:rel-1}) and (\ref{eq:rel-2}) we obtain
\begin{equation}
\begin{split}
  & 2a(2a+1) - A = - 4n(n-1) - 2n(4a+3), \\
  & - 4a^2 + 8ab_1 + 4ap - 2b_1 - p - \epsilon = 2n(n-1) + \tfrac{8}{3} n (2a-2b_1+1-p) + C_{1,n}, \\
  & B_1 - 4b_1(2a-b_1-1-p) + 4b_2(4a-3) = \tfrac{8}{3} n(b_1-2b_2) + C_{2,n},
\end{split}
\end{equation}
where, from (\ref{eq:C}), the integration constants $C_{1,n}$ and $C_{2,n}$ are given by
\begin{equation}
\begin{split}
  C_{1,n} &= - 2[4(n-1) + 4a+3] m_{(1, \dot{0})} + 2n(n-1) + \tfrac{4}{3} n(2a-2b_1+1-p), \\
  C_{2,n} &= - 2[4(n-1) + 4a+3] m_{(2, \dot{0})} + 4[2(n-1) +2a-2b_1+1-p] m_{(1,\dot{0})} \\
  & \quad - 8 m_{(1^2,\dot{0})} + \tfrac{16}{3} n(b_1-2b_2).
\end{split}
\end{equation}
On the other hand, the direct application of the FBA relations (\ref{eq:rel-1-bis}) and (\ref{eq:rel-2-bis}) yields the equivalent results
\begin{equation}
\begin{split}
  & 2a(2a+1) - A = - 4n(n-1) - 2n(4a+3), \\
  & - 4a^2 + 8ab_1 + 4ap - 2b_1 - p - \epsilon = 4n(n-1) + 4n (2a-2b_1+1-p) \\  
  & \quad - 8(n-1) m_{(1,\dot{0})} - 2(4a+3) m_{(1,\dot{0})}, \\
  & B_1 - 4b_1(2a-b_1-1-p) + 4b_2(4a-3) = 8n(b_1-2b_2) + 8(n-1) m_{(1,\dot{0})} \\
  & \quad - 8[(n-1) m_{(2,\dot{0})} + m_{(1^2,\dot{0})}] + 4 (2a-2b_1+1-p) m_{(1,\dot{0})} - 2(4a+3)
       m_{(2,\dot{0})}.
\end{split}
\end{equation}
\par
%
%
In both cases, on setting $n=0$ for instance, we get that the Schr\"odinger equation (\ref{eq:NLHO}) with $m=2$,
\begin{equation}
  A = 2a(2a+1), \qquad B_1 = 4b_1 (2a-b_1-1-p) - 4b_2 (4a-3),
\end{equation}
and $B_2$, $B_3$, $B_4$ given in (\ref{eq:B}), has an eigenvalue
\begin{equation}
  E_{0,p} = \lambda (8ab_1 + 4ap + 2a - 2b_1 - p)
\end{equation}
with the corresponding eigenfunction
\begin{equation}
  \psi_{0,p}(x) \propto x^p (1+\lambda x^2)^{-a} e^{-\lambda (b_1+2b_2) x^2 - \lambda^2 b_2 x^4}.
\end{equation}
The latter is normalizable with respect to the measure $d\mu$ provided $b_2 > 0$ if $\lambda > 0$ or $a < \frac{1}{4}$ if $\lambda < 0$ and it corresponds to a ground state for $p=0$ and to a first excited state for $p=1$.\par
%
%
{}Furthermore, for $n=1$, on taking into account that $m_{(1,\dot{0})} = z_1$, $m_{(2,\dot{0})} = z_1^2$, and $m_{(1^2,\dot{0})} = 0$ in terms of the single root $z_1$ of $y_1(z)$, we obtain that equation (\ref{eq:NLHO}) with $m=2$,
\begin{align}
  A &= (2a+2)(2a+3), \\
  B_1 &= - 2(4a+3) z_1^2 + 4(2a-2b_1+1-p) z_1 + 4b_1 (2a-b_1+1-p) - 4b_2 (4a+1),
\end{align}
and $B_2$, $B_3$, $B_4$ given in (\ref{eq:B}), has an eigenvalue
\begin{equation}
  E_{1,p} = \lambda [8ab_1 + 4ap + 2a + 6b_1 + 3p + 2 + 2(4a+3) z_1]
\end{equation}
with the corresponding eigenfunction
\begin{equation}
  \psi_{1,p}(x) \propto x^p (1+\lambda x^2)^{-a-1} [1 - z_1(1+\lambda x^2)] e^{-\lambda (b_1+2b_2) x^2 - 
  \lambda^2 b_2 x^4}.
\end{equation}
The normalizability condition is now $b_2>0$ if $\lambda>0$ or $a<-\frac{3}{4}$ if $\lambda<0$. Here $z_1$ is a real solution of the cubic equation
\begin{equation}
  (4a+3) z_1^3 - 2(2a-2b_1+1-p) z_1^2 - 4(b_1-2b_2) z_1 - 8b_2 = 0,
\end{equation}
hence, for instance,
\begin{equation}
\begin{split}
  z_1 &= \frac{2(2a-2b_1+1-p)}{3(4a+3)} + \left(- \frac{v}{2} + \sqrt{\left(\frac{v}{2}\right)^2 + 
    \left(\frac{u}{3}\right)^3}\right)^{1/3} \\
  &\quad {} + \left(- \frac{v}{2} - \sqrt{\left(\frac{v}{2}\right)^2 + \left(\frac{u}{3}\right)^3}\right)^{1/3}, \\
  u &= \frac{4}{4a+3} \left(- b_1 + 2b_2 - \frac{(2a-2b_1+1-p)^2}{3(4a+3)}\right), \\
  v &= - \frac{8}{4a+3} \biggl(b_2 + \frac{2}{27} \frac{(2a-2b_1+1-p)^3}{(4a+3)^2} \\
  &\quad {} + \frac{1}{3} \frac{(2a-2b_1+1-p)(b_1-2b_2)}{4a+3}\biggr).
\end{split}
\end{equation}
According to its value, the wavefunction $\psi_{1,p}(x)$ may correspond to a ground or second-excited state for $p=0$ and to a first- or third-excited state for $p=1$.
%
%
\section{CONCLUSION}

In the present paper, we have reconsidered the general conditions that have to be satisfied by the coefficients of a second-order  differential equation with at most $k+1$ singular points in order that the equation has particular solutions that are $n$th-degree polynomials $y_n(z)$ and we have expressed them in terms of symmetric polynomials in the polynomial solution roots. This has been done in two different ways.\par
%
%
In the first one, we have shown that these conditions involve $k-2$ integration constants, which satisfy a system of linear equations whose coefficients can be expressed in terms of elementary symmetric polynomials in the polynomial solution roots  whenever such roots are all real and distinct. In the second approach, we have considered the solution of the FBA method in its most general form under the same assumption.\par
%
%
Comparing the outcomes of both descriptions, we have proved that the above-mentioned $k-2$ integration constants can be expressed as linear combinations of monomial symmetric polynomials in the polynomial solution roots, corresponding to partitions into no more than two parts. As far as the author knows, this property  and the general solution of the FBA method in terms of such symmetric polynomials are new results.\par
%
%
In addition, their practical usefulness has been demonstrated by solving a QES extension of the Mathews-Lakshmanan nonlinear oscillator, corresponding to $k=4$.\par
%
%
\section*{APPENDIX}

\renewcommand{\theequation}{A.\arabic{equation}}
\setcounter{equation}{0}

The purpose of this appendix is to solve equation (\ref{eq:syst-C}) for $r=3$, 4, 5 and to show that the resulting expressions of $C_{1,n}$, $C_{2,n}$, and $C_{3,n}$ agree with equation (\ref{eq:C}).\par
%
%
{}For $r=3$, equation (\ref{eq:syst-C}) directly leads to
\begin{equation}
  \frac{C_{1,n}}{(k-3)!} = - [2(n-1) a_k + b_{k-1}] e_1 - \frac{2n(n-1)}{k} a_{k-1} - \frac{n}{k-1} b_{k-2},
  \label{eq:A1}
\end{equation}
which corresponds to equation (\ref{eq:C}) for $q=1$ because $e_1 = m_{(1,\dot{0})}$.\par
%
%
{}For $r=4$, equation (\ref{eq:syst-C}) becomes
\begin{align}
  & \frac{C_{2,n}}{(k-4)!} - \frac{C_{1,n}}{(k-3)!} e_1 \nonumber \\
  &= 2 [(2n-3) a_k + b_{k-1}] e_2 + \Bigl[\frac{2}{k}(n-1)(n-k) a_{k-1} + \frac{1}{k-1}(n-k+1) b_{k-2}\Bigr] e_1
       \nonumber \\
  &\quad - \frac{2n(n-1)}{k(k-1)} (2k-3) a_{k-2} - \frac{2n}{k-1} b_{k-3}.  \label{eq:A2}
\end{align}
On inserting (\ref{eq:A1}) in (\ref{eq:A2}) and using the identities $e_2 = m_{(1^2,\dot{0})}$, $e_1^2 = m_{(2,\dot{0})} + 2 m_{(1^2,\dot{0})}$, we get
\begin{align}
  \frac{C_{2,n}}{(k-4)!} &= - [2(n-1) a_k + b_{k-1}] m_{(2,\dot{0})} - [2(n-1) a_{k-1} + b_{k-2}] m_{(1,\dot{0})}
      \nonumber \\
  &\quad - 2a_k m_{(1^2,\dot{0})} - \frac{2n(n-1)}{k(k-1)} (2k-3) a_{k-2} - \frac{2n}{k-1} b_{k-3},  \label{eq:A3}
\end{align}
which agrees with equation (\ref{eq:C}) for $q=2$.\par
%
%
On setting now $r=5$ in equation (\ref{eq:syst-C}), we obtain
\begin{align}
  & \frac{C_{3,n}}{(k-5)!} - \frac{C_{2,n}}{(k-4)!} e_1 + \frac{C_{1,n}}{(k-3)!} e_2 \nonumber \\
  &= - 3[2(n-2) a_k + b_{k-1}] e_3 - \Bigl\{\frac{2}{k} [n^2 - (2k+1)n + 3k] a_{k-1} + \frac{1}{k-1} (n-2k+2) b_{k-2}
       \Bigr\} e_2 \nonumber \\
  &\quad + \Bigl\{\frac{2}{k(k-1)} (n-1)[(2k-3)n - k(k-1)] a_{k-2} + \frac{1}{k-1} (2n-k+1) b_{k-3}\Bigr\} e_1
       \nonumber \\
  &\quad - \frac{6n(n-1)}{k(k-1)} (k-2) a_{k-3} - \frac{3n}{k-1} b_{k-4}.  \label{eq:A4}
\end{align}
Here, let us employ equations (\ref{eq:A1}) and (\ref{eq:A3}), as well as the identities $e_3 = m_{(1^3,\dot{0})}$, $m_{(2,\dot{0})} m_{(1,\dot{0})} = m_{(3,\dot{0})} + m_{(2,1,\dot{0})}$, and $m_{(1^2,\dot{0})} m_{(1,\dot{0})} = m_{(2,1,\dot{0})} + 3 m_{(1^3,\dot{0})}$. For the coefficient of $m_{(1^3,\dot{0})}$ in $C_{3,n}/(k-5)!$, we obtain $- 3[2(n-2) a_k + b_{k-1}]$ from the right-hand side of (\ref{eq:A4}), $-6a_k$ from $C_{2,n} e_1/(k-4)!$, and $3[2(n-1) a_k + b_{k-1}]$ from $- C_{1,n} e_2/(k-3)!$, respectively. We conclude that $m_{(1^3,\dot{0})}$ does not occur in $C_{3,n}/(k-5)!$, which is given by
\begin{align}
  \frac{C_{3,n}}{(k-5)!} &= - [2(n-1) a_k + b_{k-1}] m_{(3,\dot{0})} - [2(n-1) a_{k-1} + b_{k-2}] m_{(2,\dot{0})}
       \nonumber \\
  &\quad - [2(n-1) a_{k-2} + b_{k-3}] m_{(1,\dot{0})} - 2a_k m_{(2,1,\dot{0})} - 2a_{k-1} m_{(1^2,\dot{0})}
       \nonumber \\
  &\quad - \frac{6n(n-1)}{k(k-1)} (k-2) a_{k-3} - \frac{3n}{k-1} b_{k-4},
\end{align}
in agreement with equation (\ref{eq:C}) for $q=3$.\par
%
%

\end{document}